# Possibility of obtaining two orders of magnitude larger electrokinetic streaming potentials, through liquid infiltrated surfaces

B. Fan and P.R. Bandaru

Mechanical and Aerospace Engineering, University of California San Diego

Material Science Engineering, University of California San Diego

**Abstract**

A methodology for substantially increasing the magnitude of the electrokinetic streaming potential ($V_s$) from ~ 0.02 V to as large as ~ 1.6 V is proposed. This is done through deploying textured, liquid-filled surfaces (*LFS*), filled with low viscosity oils, for electrolyte flow. The charge density at the electrolyte-oil interface as well as the enhanced slip may be responsible for enhancement of the $V_s$. It was found, through experimental analysis as well as computational simulations, that the fluid slip length was inversely proportional to the filling oil viscosity, and influences the $V_s$. The study provides new perspectives related to complex electrolyte flow conditions as may be relevant for energy harvesting applications.

*Introduction.* Electrokinetic flows, considering the movement of electrolyte relative to another charged surface [1,2] are relevant for understanding the effects of charge accumulation and dispersion with applications ranging from electrical power generation [3–5] to biochemical separations [6,7]. There are two major related phenomena in such flows, under a pressure difference across the microchannel, *i.e.,* (i) where the motion of ions in the electrical double layer (EDL) near a charged surface generates an electrical streaming current ($I_s$), and (ii) under open-electrical circuit conditions – where a potential difference, termed a streaming potential ($V_s$) is measured, due to charge separation. The $V_s$ may be particularly enhanced in micro- and nano-scale fluidics [3,8,9], *e.g.,* through the overlap of EDLs in nanometer size pores/channels which may enable unipolar flow and battery-like voltage sources.

Fluid flows over hydrodynamically smooth surfaces with concomitant no-slip conditions, yield low streaming currents and potential. It has then been indicated that enhanced electroosmotic mobility: $M$, may be obtained through the use of patterned [10,11] or superhydrophobic (SH) surfaces [12], in both laminar and turbulent flows. Relevant to SH surfaces [12] is an enhanced ion mobility [13], $M = \frac{\varepsilon \zeta}{\eta_e}$, where $\varepsilon$ (= $\varepsilon_o \varepsilon_r$, with $\varepsilon_o$=8.854·10$^{-12}$ C$^2$/Nm$^2$ is the vacuum permittivity and $\varepsilon_r$ is the relative permittivity) is the dielectric permittivity of the electrolyte, $\zeta$ is the zeta potential, and $\eta_e$ the electrolyte viscosity, predicated on the requirement [14–16] that the surfaces ensuring fluid slip have a significant charge density with a similar magnitude and sign as that of the no-slip surface. Traditionally, SH surfaces have been constituted through roughness on the fluid slipping surface [17–19] or through lithographic patterning [20,21], which in both cases exploits air *in* the surface to promote slip. In the context of electrokinetic flows, air would not be useful as it was conclusively determined that only a charged liquid-air interface could enhance the $V_s$ [14–16]. There is considerable ambiguity, as to whether charge exists on the air-electrolyte interface, *e.g.,* due to residual OH$^-$ ions [16]. Consequently, when the slipping surface is uncharged/partially charged, the magnitude of the $V_s$ could be diminished compared to a homogeneously charged smooth surface [14,15], as was also confirmed through our previous experimental work [22].

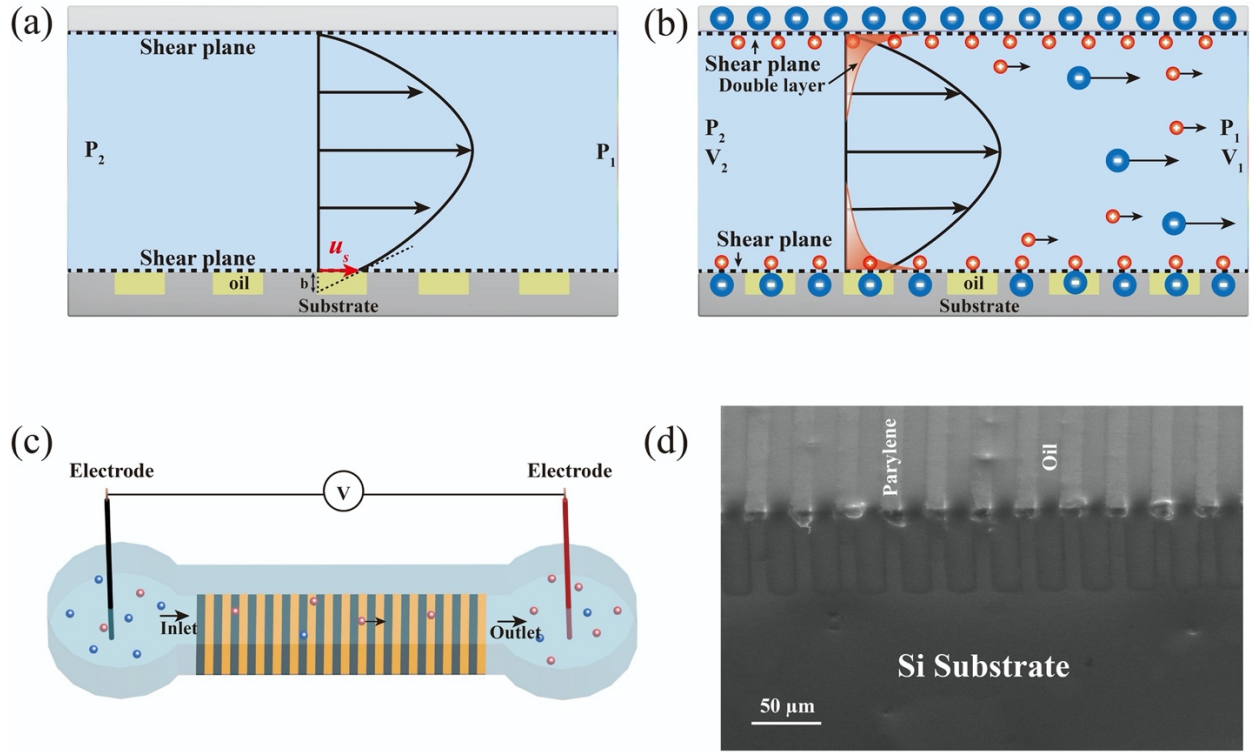

**Figure 1.** Electrokinetic flows through Poiseuille type electrolyte flow (under a pressure difference $\Delta P = P_1 - P_2$) on liquid filled surfaces (*LFS*). In addition to **(a)** finite slip velocity ($u_s$) at the interface, **(b)** a charged electrolyte-slipping surface interface ensures an enhanced streaming potential ($V_s = V_1 - V_2$). **(c)** Experimental measurement of the $V_s$ (related to the measured voltage difference: $V$, across the electrodes at the end) ensuing from pressure driven flow of salt water, **(d)** A scanning electron microscopy (SEM) image of an *LFS* filled by GPL oil.

As an alternative to conventional air-filled surfaces, liquid filled surfaces (*LFS*), fabricated by filling oil into the interstices of rectangular patterned surfaces, were studied in this work. The oil spillage out of the grooves in the pattern was considered negligible due to the enhanced surface tension forces, and was verified through microscopy [23]. The frictional drag [24–26] between the flowing electrolyte and a solid surface could also be diminished. The advantages of an *LFS*, constituted from liquid as well as the intervening hydrophobic solid surface, are that a definitive charge density would be ensured at the slipping surface in addition to a finite slip velocity [27]. **Fig. 1 (a)** emphasizes the flow profile (incorporating the slip velocity: $u_s$, as well as the associated slip length: $b$) [22] while **Fig. 1(b)** depicts the associated electrokinetic flows, over the *LFS*.

*Theory*. The electrokinetic flow of an electrolyte, under a pressure difference ($\Delta P$), has been traditionally considered through the Helmholtz-Smoluchowski (H-S) relation, which on homogeneous surfaces is of the form [28,29]: $V_s = \frac{M}{\sigma} \Delta P$, with $\sigma$ as the electrolyte conductivity. Further assumptions [30] underlying the relation include negligible surface (/substrate) conductivity and a very small EDL thickness [31], with Poiseuille flow of the electrolyte. Moreover, there is an implicit assumption of the no-slip boundary condition [32], with a finite flow velocity only at a certain distance away (corresponding to a shear plane) into the electrolyte. The $\zeta$ would correspond to the electrical potential at the edge of the shear plane [29,30]. Most work on harnessing the $V_s$, to date, has indeed been concerned with flows over smooth surfaces (where the scale of roughness is smaller than the Debye length: $\lambda_D$), and consequently the use of the H-S relation implies mV levels of the measured $V_s$ [28], *e.g.,* with a $\Delta P \sim 1000$ Pa, and 0.1 mM L$^{-1}$ NaCl, with $\varepsilon_r \sim 80$, $\eta_e \sim 10^{-3}$ Pa·s, $\zeta \sim 25$ mV, $\sigma \sim 10^{-3}$ S m$^{-1}$, a $V_s$ of $\sim 18$ mV.

However, it was previously shown [22] that a larger than two-fold increase in the $V_s$ may be obtained through the use of the *LFS* with specific oils. While a finite charge density at the electrolyte-oil interface is one possible reason for the increase, the rationale for the choice of the filling liquid, say, with respect to the fluid slip has been unclear. Here, we discuss specific correlations between the viscosity of the filling liquid in the *LFS* with the experimentally obtained $V_s$. We aim to provide a deeper understanding and new perspectives on electrokinetic flows through such investigations.

It has previously been considered, based on molecular dynamics (MD)-based simulations [33] that slip may mobilize the Stern layer, significantly enhancing the $\zeta$. Due to the slip, the shear plane would be moved closer to the substrate. As the surface electrical potential is reduced, *e.g.,* exponentially, away from the surface, a proximate shear plane may yield a larger $\zeta$ with concomitantly increased $V_s$. Here, the zero velocity boundary condition at the surface (*y=0*) would be replaced with a Navier slip condition: $u_s(y=0) = b \frac{\partial u\ (y=0)}{\partial y})$ with $b$ as the slip length [34,35], and $u_s$ as the slip velocity: **Fig. 1 (a)**. The $V_s$ will be enhanced, over that predicted through the H-S relation, by a factor $\frac{b_{eff}}{\lambda_D}$, where $b_{eff}$ is the effective slip length – an average local slip length [28], *i.e.,*

$$V_s = \frac{\varepsilon \zeta}{\eta_e \sigma} \Delta P \left(1 + \frac{b_{eff}}{\lambda_D}\right) \tag{1}$$

For a rectangular groove patterned surface, the $b_{eff}$ in the direction parallel (*i.e.,* $b_{eff}^{\parallel}$), and that perpendicular ($b_{eff}^{\perp}$) to the grooves [36,37], may be estimated through the following relations:

$$b_{eff}^{\parallel} = \frac{L}{\pi} \frac{\ln[\sec(\frac{\pi\phi}{2})]}{1+\frac{L}{\pi b}\ln[\sec(\frac{\pi\phi}{2})+\tan(\frac{\pi\phi}{2})]} \tag{2a}$$

$$b_{eff}^{\perp} = \frac{L}{2\pi} \frac{\ln[\sec(\frac{\pi\phi}{2})]}{1+\frac{L}{2\pi b}\ln[\sec(\frac{\pi\phi}{2})+\tan(\frac{\pi\phi}{2})]} \tag{2b}$$

$L(= w + d)$, is the groove pattern period with $w$ as the groove width and $d$ as the lateral length of the solid surface, $b = w \frac{\eta_e}{\eta_{oil}} \beta$, is the local constant slip length, and $\eta_{oil}$ is the viscosity of the liquid (/oil) in the grooves [36]. Generally, $\beta$ has been modeled with different values for parallel and transverse grooves [36], *i.e.*, $\beta_\| = \frac{\text{erf}(\frac{q_x h}{w})}{q_x}$, and $\beta_\perp = \frac{\text{erf}(\frac{q_y h}{w})}{4 q_y}$, where $h$ is the groove depth, $q_x \simeq 3.1$, $q_y \simeq 2.17$. When $h/w \gg 1$, as in our case, with $h \sim 100$ μm and $w \sim 18$ μm, the $\beta_\|$ and $\beta_\perp$ may be estimated to be $\sim 0.32$ and $\sim 0.12$, respectively. Consolidating the above relationships, we obtain:

$$V_s = \frac{\varepsilon \zeta}{\eta_e \sigma} \Delta P \left(1 + \frac{m}{1 + n \eta_{oil}}\right) \quad (3)$$

Here, $m$ and $n$ are two groove geometry-dependent parameters, *i.e.,* for parallel grooves:

$$m_\| = \frac{L}{\pi} \frac{\ln[\sec(\frac{\pi \phi}{2})]}{\lambda_D}, \quad n_\| = \frac{L}{\pi} \frac{\ln[\sec(\frac{\pi \phi}{2}) + \tan(\frac{\pi \phi}{2})]}{w \beta \eta_e} \quad (4a)$$

While for transverse grooves:

$$m_\perp = \frac{L}{2\pi} \frac{\ln[\sec(\frac{\pi \phi}{2})]}{\lambda_D}, \quad n_\perp = \frac{L}{2\pi} \frac{\ln[\sec(\frac{\pi \phi}{2}) + \tan(\frac{\pi \phi}{2})]}{w \beta \eta_e} \quad (4b)$$

The aim is to experimentally verify Eqn. (3) with an explicit consideration of the nature of the *LFS,* to yield insights into the influence of surfaces on electrokinetic behavior.

*Experiment*. The $V_s$ was monitored in a microfluidics-based setup, with a microchannel ($\sim 250$ μm in height, 11.8 cm in length and 0.9 cm in width) using salt water (with NaCl dissolved in water of varying concentrations from 0.1 mM to 100 mM) under pressure driven Poiseuille flow: **Fig. 1(c).** The channel surfaces were constituted from an upper surface (silicone coated onto polycarbonate) and a bottom test surface – which was of the *LFS* type. The *LFS* was fabricated by infiltrating a series of GPL oils with systematically varying viscosities into lithographically patterned channels. The bare channel fabrication and related details have been previously discussed [22,23]. Here, we report on the obtained results from the *LFS* constituted from a given groove width ($w = 18$ μm), and groove period: $L (= w + d) = 36$ μm. We define a groove fraction, $\phi = w/L$, to characterize the patterned surfaces. The $\eta_{oil}$ was varied over two orders of magnitude (in the range of 30 mPa·s to 3000 mPa·s), as indicated in **Table I,** for probing the electrokinetic potentials. The oils were found to be immiscible with the aqueous electrolyte (NaCl in deionized water) and with low surface energy (Dupont Krytox GPL). The GPL (General Purpose Lubricant) represents a family of widely used perfluorinated oils known to be inert and stable over a wide temperature range. The top and bottom surfaces were separated by silicone rubber spacer to adjust the height of the microchannel, and Ag/AgCl electrodes were immersed in the reservoirs at the either end to measure the potential difference, for the $V_s$. The pressure drop ($\Delta P$) along the channel length, in the range of 200 Pa to 1200 Pa, was measured by a manometer (UEI EM152) and checked to be in correspondence with the Poiseuille flow. The chosen range of pressure yielded stable and reproducible $V_s$. The experiments were performed with the flow direction perpendicular to the grooves considering the robustness of the *LFS*, *i.e.,* whether the filling liquid would experience drainage under external shear flow [38]. It was discussed previously that the filling liquid could be retained indefinitely in the grooves in such a configuration [39] and where the period ($L$) was less than a critical length [38]: $L_\infty$, inversely proportional to the $w/h$ ratio. Since, for the given conditions, $L_\infty$ was estimated to be of the order of millimeters, it was assumed, from such consideration as well from SEM observations after the experiments, **Fig. 1(d)**, that the *LFS* are stable. The $V_s$ was measured six times at each applied pressure and the average value was used.

**Table I.** Viscosity of the oils deployed in the *LFS*

| Oils used in *LFS* | $\eta_{oil}$ (*mPa.s*) |
|---|---|
| GPL 101 | 33 |
| GPL 102 | 73 |
| GPL 104 | 341 |
| GPL 105 | 1012 |
| GPL 107 | 2993 |

*Results and Discussion:* The measured $V_s$ as a function of the $\Delta P$ are in general accord with the Helmholtz-Smoluchowski (H-S) relation, through the obtained linear variations for different *LFS* **Fig. 2 (a)**. For 0.1 mM NaCl solution, with a Debye length $\lambda_D \sim 30$ nm, we estimate, from Eqn. 4(b), $m_\perp = 66.2$, and $n_\perp = 2.3$. The $\zeta$ of *LFS* was estimated to be $\sim 30.9$ mV. Inserting the numerically obtained values into Eqn. (3) with $\Delta P = 1200$ Pa, we obtain the following relation between the $V_s$ and $\eta_{oil}$ for the *LFS* as:

$$V_s(mV) = 24.15\left(1 + \frac{66.19}{1+2.34*\eta_{oil}\,(cP)}\right) \quad (5)$$

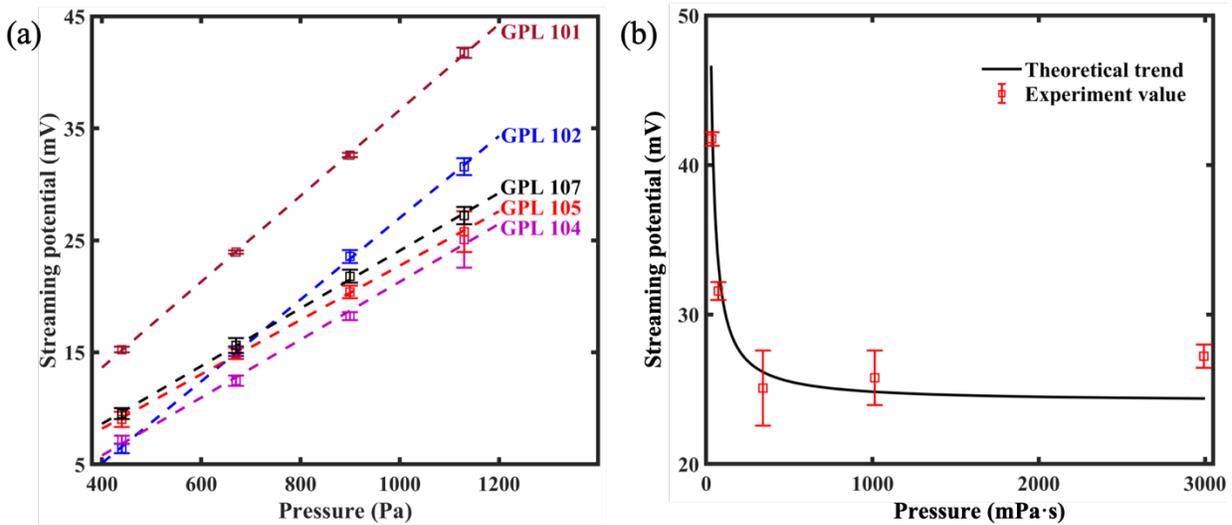

**Figure 2.** (a) The measured streaming potential ($V_s$) on liquid filled surface (*LFS*) using a series of GPL oils, of varying viscosity, with 0.1 mM NaCl as the electrolyte solution scales linearly with the applied pressure drop, (b) A comparison of the experimentally measured $V_s$ (when the pressure drop: $\Delta P = 1200\ Pa$) to the analytical relationship of Eqn. (3).

A comparison of the experimentally obtained $V_s$ with the analytically evaluated Eqn. (5) is indicated in **Fig. 2(b)** and shows excellent agreement ($R^2 = 0.99$). Based on such agreement, we predict an even larger $V_s$ with smaller $\eta_{oil}$. Indeed, using aqueous media (with $\eta \sim 1$ mPa·s) or hydrocarbon-based liquids (with $\eta \sim 0.2$ mPa·s) would yield $V_s$ values of the order of 0.5 V and 1.1 V, respectively. The relationship also allows us to predict an upper limit to the $V_s$, obtained when $\eta$ tends to zero, of ~ 1.6 V, approaching the potential difference of batteries.

We used finite element methodologies (FEM), through the incompressible Navier-Stokes equations, to model the fluid flow over *LFS* with $\eta_{oil}$ in the groove over the range of values in **Table 1**. When electrolyte flows above the *LFS*, the shear stress at the electrolyte - oil interface is implicated in the generation of vortices as indicated in **Fig. 3 (a).** The Poiseuille flow of the electrolyte in the channel was considered with non-zero slip velocity ($u_s$) [40] at the interface, as in **Fig. 3 (b)**. The $u_s$ as well as the slip length was found to be inversely proportional to $\eta_{oil}$ as indicated in **Fig. 3(c)**. The related shear rate was estimated from the velocity profile, and a corresponding fluid slip length ($b_{sim}$) was obtained from the Navier slip boundary condition. The $b_{sim}$ was compared to the theoretical estimate, *i.e.,* with $b_{theo}$ $(= w \frac{\eta_e}{\eta_{oil}} \beta)$, in **Table II**, and close correspondence was seen.

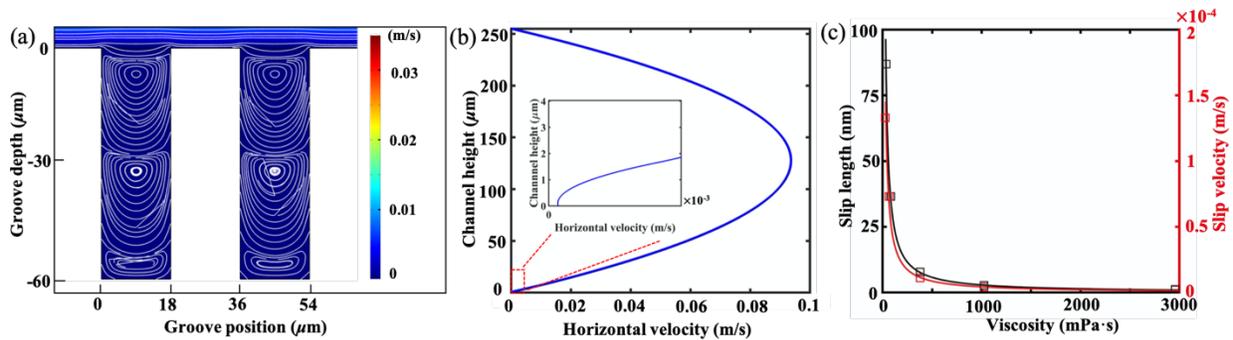

**Figure 3. (a)** Simulated flow velocity near the grooves for oil (GPL 101) filled liquid filled surfaces (*LFS*) with groove height $h = 60$ $\mu m$, groove width $w = 18$ $\mu m$, **(b)** The velocity profile between the liquid-oil interface (*bottom*) and the upper surface of the microchannel; the *inset* indicates the non-zero slip velocity at the liquid-oil interface, **(c)** The simulated slip length (*left axis*) as well as the slip velocity (*right axis*) scales inversely as the oil viscosity (at a $\Delta P = 1200$ Pa).

**Table II.** The electrolyte fluid slip length ($b$), from simulations: $b_{sim}$ in comparison with the theoretical estimates ($b_{theo}$).

| Oils | $b_{sim}$ (nm) | $b_{theo}$ (nm) |
|---|---|---|
| GPL 101 | 87.0 | 65.0 |
| GPL 102 | 40.9 | 30.0 |
| GPL 104 | 8.7 | 6.3 |
| GPL 105 | 3.0 | 2.1 |
| GPL 107 | 1.0 | 0.7 |

*Conclusions.* The proposed work has indicated that streaming potentials, as large as 1.6 V, may possibly be obtained through the use of specifically structured surfaces, such as the *LFS,* in comparison to the typical values of ~ 0.02 V using smooth surfaces or even conventional superhydrophobic surfaces. It was shown that the viscosity of the infiltrating oil in the *LFS* is critical to the obtained $V_s$. An analytically derived relationship was confirmed experimentally and has been used to predict the limits of the $V_s$. The *LFS,* in addition to enhancing the fluid slip velocity and slip length, provides a charged electrolyte-oil interface which may contribute to the plausible orders of magnitude enhancement of the $V_s$. While it was previously indicated that [9,41] that overlap of the EDLs in nanoscale channels may be necessary for boosting the magnitude of the electrokinetic potential, flow and throughput restrictions are a major constraint [42]. The presented work provides an alternative for achieving large $V_s$ at the *micro*scale. With such plausibility, our results provide much motivation for aiming at more detailed understanding of electrokinetics on hybrid/non-homogeneous surfaces and open new perspectives for guiding multiphase flow and related biology and energy harvesting applications.

We gratefully acknowledge financial support from the National Science Foundation (NSF: CBET 1606192). We thank Dr. S. Rubin for valuable discussions.

**Corresponding Author:** P.R. Bandaru, email: pbandaru@ucsd.edu